\documentclass[12pt]{iopart}

\usepackage{epsfig,graphics}
\begin{document}
\title {Nuclear inputs of key iron isotopes for core-collapse modeling and simulation}
\author{Jameel-Un Nabi$^1,^2$ \footnote{Corresponding author} and Abdel Nasser Tawfik$^2$}
\address{1 Faculty of Engineering Sciences, GIK Institute of Engineering Sciences and
Technology, Topi 23640, Khyber Pakhtunkhwa, Pakistan
\\ 2  Egyptian Center for Theoretical
Physics, MTI Modern University, Al-Mokattam, Cairo, Egypt}
\eads{jameel@giki.edu.pk, a.tawfik@mti.edu.eg}

\begin{abstract}
From the modeling and simulation results of presupernova evolution
of massive stars, it was found that isotopes of iron,
$^{54,55,56}$Fe, play a significant role inside the stellar cores,
primarily decreasing the electron-to-baryon ratio ($Y_{e}$) mainly
via electron capture processes thereby reducing the pressure
support. The neutrinos produced, as a result of these capture
processes, are transparent to the stellar matter and assist in
cooling the core thereby reducing the entropy. The structure of the
presupernova star is altered both by the changes in $Y_{e}$ and the
entropy of the core material. Here we present the microscopic
calculation of Gamow-Teller strength distributions for isotopes of
iron. The calculation is also compared with other theoretical models
and experimental data. Presented also are stellar electron capture
rates and associated neutrino cooling rates, due to isotopes of
iron, in a form suitable for simulation and modeling codes. It is
hoped that the nuclear inputs presented here should assist
core-collapse simulators  in the process of fine-tuning of the
$Y_{e}$ parameter during various phases of presupernova evolution of
massive stars. A reliable and accurate time evolution of this
parameter is a possible key to generate a successful explosion in
modeling of core-collapse supernovae.
\end{abstract}
 \maketitle


\section{Introduction}
The understanding of production of energy and matter in stars have
come a long way since the seminal paper by Bethe \cite{Bet39}.
Almost half a century later Bethe again tried to summarize the
mechanisms of supernova explosions \cite{Bet90}. Today we know that
there are two key quantities that control the complex dynamics of
core-collapse supernovae: the lepton-to-baryon fraction ($Y_{e}$)
and the core entropy. It is the weak-interaction rates (mainly
electron capture and $\beta$-decay) that alter $Y_{e}$ and hence
control the time evolution of this key parameter. At the same time
the (anti)neutrinos produced by these weak rates are transparent to
stellar matter for densities less than around 10$^{11}$ gcm$^{-3}$
and helps in removing energy and entropy from the stellar core.

Nuclei in the mass range A $\sim$ 60, at stellar densities less than
around 10$^{11}$ gcm$^{-3}$, posses electron chemical potential of
the same order of magnitude as the nuclear Q-value. Under such
conditions the electron capture rates are sensitive to the detailed
Gamow-Teller (GT) distributions. A reliable and microscopic
calculation of ground and excited states GT distribution functions
is then in order. Further to this, electron capture rates and
associated neutrino cooling rates (as a function of stellar
temperature, density and Fermi energy) are required by core-collapse
simulators for modeling the dynamics and supernova collapse phase of
massive stars (e.g. see the review article by Strother and Bauer
\cite{Str09}).

Isotopes of iron (mainly $^{54,55,56}$Fe) play an effective role in
the presupernova evolution of massive stars (see for example the
simulation results of Refs. \cite{Heg01,Auf94}). The microscopic
calculation of weak rates of these isotopes of iron was calculated
by Nabi \cite{Nab09}. Later a detailed analysis of stellar electron
capture rates \cite{Nab11} and neutrino cooling rates \cite{Nab11a}
on these iron isotopes were presented by the same author (see also
\cite{Nab12,Nab13}). Keeping in view the paramount importance of
weak rates of these iron isotopes in modeling and simulation of
core-collapse supernovae, we present, in this paper, a detailed
analysis of GT strength distributions as well as electron capture
and neutrino cooling rates due to $^{54,55,56}$Fe suitable for
incorporation in stellar codes. The pn-QRPA model, first developed
by Halbleib and Sorensen \cite{Hal67}, and extended to nuclear
excited states by Muto and collaborators \cite{Mut92}, was used to
calculate the ground and excited states GT strength distribution
functions and stellar electron capture and neutrino cooling rates in
a form suitable for use in modeling and simulation codes of
core-collapse supernovae. Wherever possible we also compare and
contrast our calculation with experiments and previous theoretical
calculations. It is expected that the present work can assist in
fine-tuning of the temporal variation of $Y_{e}$ parameter which is
a key to generation of a successful explosion in modeling and
simulation of core-collapse supernovae.

We present our calculation and results in next section and summarize
the main conclusions in Section~3.
\section{Calculations and Results}
Calculation of electron capture (ec) and neutrino cooling rates
($\nu$) in stellar matter is given by
\begin{equation}
\lambda ^{^{ec(\nu)} } _{ij} =\left[\frac{\ln 2}{D}
\right]\left[f_{ij}^{ec(\nu)} (T,\rho ,E_{f} )\right]\left[B(F)_{ij}
+\left({\raise0.7ex\hbox{$ g_{A} $}\!\mathord{\left/ {\vphantom
{g_{A}  g_{V} }} \right.
\kern-\nulldelimiterspace}\!\lower0.7ex\hbox{$ g_{V}  $}}
\right)^{2}_{eff} B(GT)_{ij} \right].
\end{equation}
The calculation of Fermi and GT reduced transition probabilities
(within the pn-QRPA formalism) and phase space integrals in Eqt.~1
are not shown here because of space limitations and can be seen from
Ref. \cite{Nab04}. A quenching factor of 0.6 was introduced in the
calculation \cite{Nab09}. The value of D was taken as 6295 s.

The total electron capture (neutrino cooling) rate per unit time per
nucleus is given by
\begin{equation}
\lambda^{ec(\nu)} =\sum _{ij}P_{i} \lambda _{ij}^{ec(\nu)},
\end{equation}
where it is to be noted that $\lambda_{ij}^{\nu}$ contains
contribution from both electron capture \textit{and} positron decay
rates for the transition $i$ (parent excited state) $\rightarrow j$
(daughter excited state) and $P_{i}$ is the probability of
occupation of parent excited state which follows the normal
Boltzmann distribution.

We next present the calculation of GT strength distribution
functions within the pn-QRPA formalism which is later used to
calculate the stellar electron capture and neutrino cooling rates.
Wherever available, we also compare and contrast our calculation
with measured data and previous theoretical calculations.

The pn-QRPA calculated GT strength values along the electron capture
direction, B(GT$_{+}$), for $^{54}$Fe are given in Table~\ref{ta1}.
We also present B(GT$_{+}$) distributions calculated using other
pn-RPA scenarios for mutual comparison in Table~\ref{ta1} (adapted
from Ref. \cite{Nab13a}). The first three columns show calculated
strength distributions using semi-realistic/semi-empirical shell
model interactions. All three interactions started from realistic
nucleon-nucleon interactions, from which an effective interaction
(e.g., a G-matrix) was derived. The three interactions are similar,
but have different starting points and were fitted to different data
sets, with the following semi-realistic/semi-empirical interactions:
the modified Kuo-Brown interaction KB3G \cite{Pov01}, the Tokyo
interaction  GXPF1 \cite{Hon02} and the Brown-Richter  interaction
FPD6 \cite{Ric91}. For further details on these interactions we
refer to Ref. \cite{Nab13a}. The pn-QRPA calculated B(GT$_{+}$)
values are given in fourth column. It is to be noted that calculated
values of B(GT$_{+}$) of magnitude less than 10$^{-3}$ are not shown
in Table~\ref{ta1} for space consideration. The pn-QRPA model
calculates GT transitions at low excitation energy as against other
three shell model interactions. It is further noted that the pn-QRPA
calculated distribution is better fragmented than other RPA models
and follows the trend of the measured data, albeit with a much
higher magnitude of strength distribution. The measured data is
shown in last column and is adapted from the $^{54}$Fe(n,p)
reaction, measured at 97 MeV for excitation energies in $^{54}$Mn
\cite{Roe93}, and follows a more or less normal distribution till
8.5 MeV in $^{54}$Mn. The fragmented data of pn-QRPA model is
attributed to an optimum choice of particle-particle and
particle-hole GT strength interaction parameters discussed later
(for details see Ref. \cite{Nab13}).

Table~\ref{ta2} shows a similar calculated B(GT$_{+}$) distribution
for $^{55}$Fe. Once again calculated values of B(GT$_{+}$) of
magnitude less than 10$^{-3}$ are not shown in Table~\ref{ta2} to
save space. No experimental data is available for comparison for
this odd-A nucleus of iron. Here one notes that the pn-QRPA model
calculates high-lying transitions in daughter nucleus. Further the
model calculates GT strength function up to around 14 MeV in
$^{55}$Mn. For $^{55}$Fe nucleus, the excited states within the pn-QRPA formalism can be constructed:\\
(1) by lifting the odd neutron from ground state to excited states (one-quasiparticle state),\\
(2) by three-neutron states, corresponding to excitation of a neutron (three-quasiparticle states), or,\\
(3) by one-neutron two-proton states, corresponding to excitation of
a proton (three-quasiparticle states). \\ The formulae for reduction
of three-quasiparticle states to correlated one-quasiparticle state
can be seen from Ref. \cite{Nab10}. The KB3G interaction saturates
first to its maximum strength and also gives the lowest values for
the centroid of the B(GT$_{+}$) distribution function.

The calculated and measured B(GT$_{+}$) distributions for $^{56}$Fe
are presented in Table~\ref{ta3}. Calculated values of B(GT$_{+}$)
of magnitude less than 10$^{-3}$ are not shown in Table~\ref{ta3}
for space consideration. One notes that, akin to the case of
$^{54}$Fe, , the pn-QRPA model calculates low-lying GT strengths (as
also measured in the experiment \cite{Elk94}). The calculated peak
around 4 MeV in the pn-QRPA model takes care of low-lying measured
data. It is easily seen from Table~\ref{ta3} that the pn-QRPA best
mimics the trend in the measured GT data and is a triumph of the
pn-QRPA model.

The key statistics of calculated GT strength distributions for
$^{54,55,56}$Fe is presented in Table~\ref{ta4}. For the sake of
completeness we present the data of GT distributions both in
electron capture and in $\beta$-decay directions. We present the
centroid E(GT), width and total GT strength for both electron
capture and $\beta$-decay directions. Centroids and widths are given
in units of MeV and B(GT) strengths are given in units such that
B(GT) = 3 for neutron decay. The centroids and widths were
calculated from the reported measured data and all measured data are
given to one decimal place in Table~\ref{ta4}. For the $\beta^{-}$
side the measured data for $^{54}$Fe were taken from Refs.
\cite{Vet89,Ada12,Rap83,And90}.  For the electron capture direction,
experimental data for $^{54}$Fe were taken from Refs.
\cite{Roe93,Vet89}. Measured data for $^{56}$Fe in the electron
capture direction were taken from Refs. \cite{Roe93,Elk94} while for
the $\beta^{-}$ direction we only quote the reported value of $\sum
B(GT_-)$ = 9.9 $\pm$ 2.4 by Rapaport and collaborators \cite{Rap83}.
The authors were unable to extract GT strengths for discrete excited
states beyond 5.9 MeV in $^{56}$Co making it impossible for us to
calculate the centroid and width in this case. It is to be noted
that we used a quenching factor of 0.6 for the calculated GT
strength using the pn-QRPA model \cite{Nab11}. Note that the shell
model interactions calculated strengths were quenched by a universal
quenching factor of 0.55 rather than 0.6 (see \cite{Nab13a}).
Table~\ref{ta4} shows that the pn-QRPA model calculates the centroid
at a much lower energy than other shell model interactions for
$^{54,56}$Fe and $^{55}$Fe (along $\beta^{-}$ direction). Further in
all cases it is seen that the pn-QRPA model best reproduces the
placement of measured centroid. The comparison is exceptionally good
for the case of GT$_{-}$ centroid of $^{54}$Fe and for the GT$_{+}$
centroids of $^{56}$Fe. On the other hand the GXPF1 interaction
calculates the highest centroid in daughter nuclei. Comparison of
pn-QRPA calculated total GT strength with measured data is also
reasonable. One should also keep in mind the uncertainties present
in measurements as well as slightly different values of quenching
factor used in pn-QRPA and shell model interactions before comparing
the calculated numbers with experimental data. Table~\ref{ta4} shows
that the pn-QRPA indeed is a good model for calculation of GT
strength distribution for isotopes of iron. The pn-QRPA calculated
GT strength distributions are in decent comparison with measurements
and should therefore translate into a reliable calculation of
stellar electron capture and neutrino cooling rates which we present
below.

\begin{figure}[htbp]
\begin{center}
\includegraphics[width=0.6\textwidth]{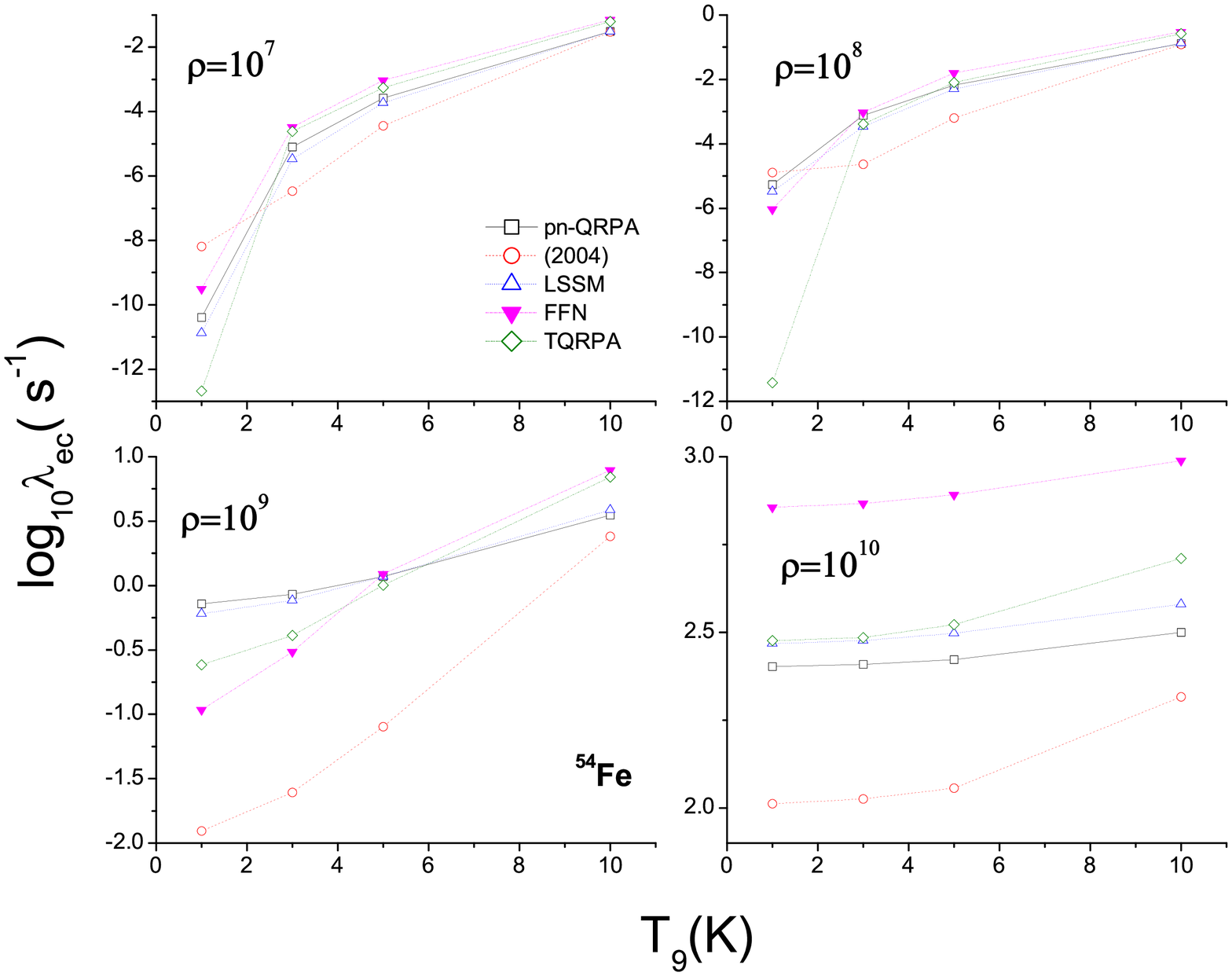}
\caption{\scriptsize Comparison of electron capture rates on
$^{54}$Fe using current pn-QRPA model (pn-QRPA), previous pn-QRPA
calculation (2004) \cite{Nab04}, shell model calculation (LSSM)
\cite{Lan00}, calculation done by Ref. \cite{Ful82} (FFN) and
calculation by Ref. \cite{Dzh10} (TQRPA) at different selected
stellar temperature and density zones. }\label{figure4}
\end{center}
\end{figure}

We first present the calculation of stellar electron capture rates
and also compare them with previous theoretical calculations.
Fig.~\ref{figure4} shows a four-panel graph depicting the electron
capture rates on  $^{54}$Fe. The upper left, upper right, lower left
and lower right panels show electron capture rates calculated at
stellar densities of 10$^{7}$, 10$^{8}$, 10$^{9}$ and 10$^{10}$
gcm$^{-3}$, respectively. T$_{9}$ gives stellar temperature in units
of 10$^{9}$ K. All electron capture rates are depicted in
logarithmic scale (to base 10) and calculated in units of $s^{-1}$.
The temperature and density zone selected for $^{54}$Fe is pertinent
for presupernova evolution of massive stars as per simulation
results of Heger and collaborators \cite{Heg01}. In each panel we
compare our results with four previous calculations: (i) previous
pn-QRPA calculation of Nabi and Klapdor-Kleingrothaus \cite{Nab04}
(2004); (ii) large scale shell model calculation \cite{Lan00}
(LSSM); (iii) the pioneer weak-interaction mediated rate
calculations performed by Fuller, Fowler and Newman \cite{Ful82}
(FFN), and (iv) the pn-QRPA model extended to finite temperature
using the thermofield dynamics formalism by Dzhioev and
collaborators \cite{Dzh10} (TQRPA). Details of other theoretical
calculations can be found in the respective references. It can be
noted from upper left panel of Fig.~\ref{figure4} that at a stellar
density of 10$^{7}$ gcm$^{-3}$ and T$_{9}$ =1, the current pn-QRPA
calculation is around 3 orders of magnitude smaller than the
previous pn-QRPA calculation of 2004 \cite{Nab04}. The two
calculations get in better agreement at higher temperatures and
densities. The difference arises because of a better choice of model
parameters in current pn-QRPA calculation. In the current pn-QRPA
model we took the values of $\chi$ (particle-hole GT strength
parameter) and $\kappa$ (particle-particle GT strength parameter) as
0.15 MeV and 0.07 MeV, respectively. The values of these crucial
parameters were different in previous pn-QRPA calculation. For a
judicious selection of these GT strength parameters we refer to
\cite{Nab13}. The deformation parameter is recently argued to be one
of the most important parameters in pn-QRPA calculations
\cite{Ste04} and as such rather than using deformations calculated
from some theoretical mass model (as used in previous pn-QRPA
calculation) the experimentally adopted value of the deformation
parameters for $^{54,56}$Fe, extracted by relating the measured
energy of the first $2^{+}$ excited state with the quadrupole
deformation, was taken from Raman et al. \cite{Ram87}. The
incorporation of experimental deformations lead to an overall
improvement in the calculation as discussed earlier in Ref.
\cite{Nab10}. Our calculation is in overall very good agreement with
LSSM results and around one order of magnitude smaller than
corresponding FFN numbers. However it is to be noted that at T$_{9}$
=1, the reported electron capture rates on $^{54}$Fe is around a
factor three bigger than LSSM rates. The TQRPA electron capture
rates are suppressed by around two orders of magnitude at T$_{9}$
=1. The comparison improves as temperature and density increase and
TQRPA rates are doubled at T$_{9}$ =10.  It appears that TQRPA rates
are too sensitive to temperature changes. Moving to upper right
panel (stellar density of 10$^{8}$ gcm$^{-3}$) one notes that all
calculations are within an order of magnitude differences except the
TQRPA results which are around six orders of magnitude smaller at
T$_{9}$ =1 than the reported rates. The two lower panels (stellar
densities of 10$^{9,10}$ gcm$^{-3}$) also depict calculated rates
which are within an order of magnitude differences of one another.
The bigger electron capture rates on $^{54}$Fe as calculated by FFN
are due to the fact that FFN did not take into effect the process of
particle emission from excited states and their parent excitation
energies extended well beyond the particle decay channel. These high
lying excited states begin to show their contribution to the total
electron capture rate specially at high temperatures and densities.
It is also to be noted that FFN neglected the quenching of the total
GT strength in calculation of their stellar electron capture rates.
Fig.~\ref{figure4} shows the fact that all calculations get in
better agreement with each other at high densities whereas
noticeable differences are seen at low temperature and densities.

\begin{figure}[htbp]
\begin{center}
\includegraphics[width=0.6\textwidth]{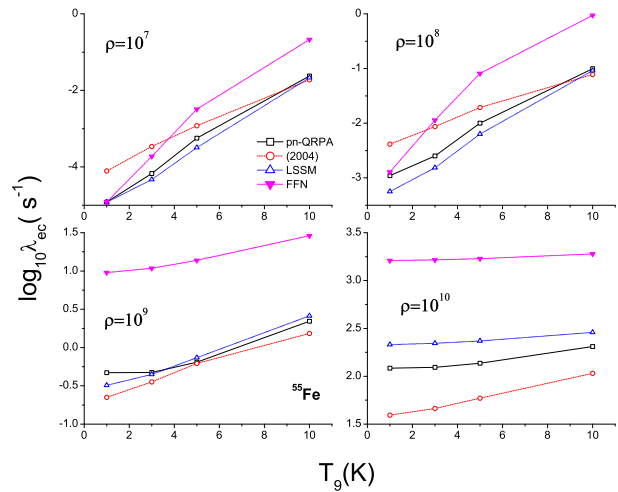}
\caption{\scriptsize Same as Figure~\ref{figure4} but for $^{55}$Fe.
TQRPA model was not used to calculate electron capture rates on
$^{55}$Fe. }\label{figure5}
\end{center}
\end{figure}

A similar comparison of electron capture rate calculations on
$^{55}$Fe is shown in Fig.~\ref{figure5}. TQRPA formalism was not
used to calculate electron capture rates on this odd-A isotope of
iron and as such not present in Fig.~\ref{figure5}. At stellar
density of around 10$^{7}$ gcm$^{-3}$  (upper left panel), one notes
that reported electron capture rates are roughly an order of
magnitude smaller as compared to previous pn-QRPA calculation. Only
at T$_{9}$ =10 do the two calculations come in agreement. The reason
for this difference was already discussed above. The comparison with
FFN calculation is in reverse order. Here FFN calculated electron
capture rates are in good agreement with reported rates at T$_{9}$
=1 and at all higher temperatures the reported rates are smaller by
an order of magnitude. Once again the comparison with LSSM rates is
excellent and speaks for the success of the pn-QRPA model. The
comparison between different theoretical models remains more or less
same at stellar density of 10$^{8}$ gcm$^{-3}$ (upper right panel).
At still higher density of 10$^{9}$ gcm$^{-3}$ (lower left panel)
the comparison of reported rates with previous pn-QRPA calculation
and LSSM is excellent whereas the FFN rates are around two orders of
magnitude bigger for reasons already mentioned above. A similar
comparison with previous calculations is also witnessed at high
stellar density of 10$^{10}$ gcm$^{-3}$ (lower right panel). For
further details we refer to Ref. \cite{Nab11}.

\begin{figure}[htbp]
\begin{center}
\includegraphics[width=0.6\textwidth]{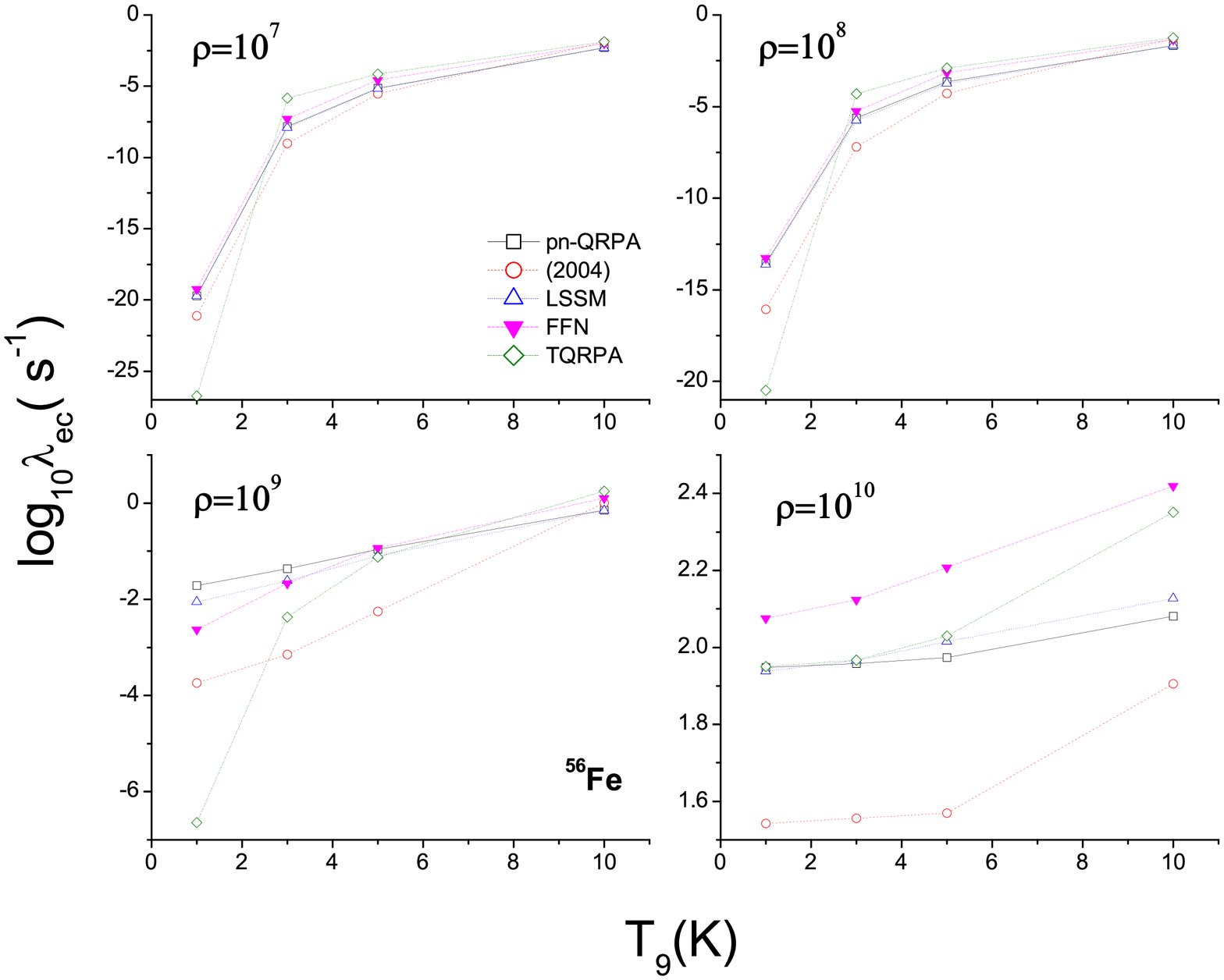}
\caption{\scriptsize Same as Figure~\ref{figure4} but for
$^{56}$Fe.}\label{figure6}
\end{center}
\end{figure}

The electron capture rate calculations on $^{56}$Fe is shown in
Fig.~\ref{figure6}. At density of around 10$^{7}$ gcm$^{-3}$  (upper
left panel)  the comparison of reported rates with LSSM rates is
perfect. The previous pn-QRPA calculation and FFN rates differ by up
to an order of magnitude with current reported electron capture
rates for reasons stated before. The TQRPA rates are suppressed by a
whopping seven orders of magnitude at low temperature of T$_{9}$ =1.
However, by merely changing T$_{9}$ =1 to T$_{9}$ =1.5, the TQRPA
rates increase by as much as up to 4 orders of magnitude bigger than
the corresponding pn-QRPA and LSSM rates. It looks that TQRPA rates
are very sensitive to temperature changes for the case of $^{56}$Fe
at low densities. The comparison improves with increasing stellar
temperatures and densities. Once again at T$_{9}$ =10 the TQRPA
rates are roughly double the pn-QRPA and LSSM rates. The upper right
panel shows the calculated rates at density of 10$^{8}$ gcm$^{-3}$.
The LSSM rates are in superb agreement whereas FFN rates are bigger
by up to an order of magnitude as compared to reported rates. The
previous pn-QRPA rates are smaller by up to two orders of magnitude
at T$_{9}$ =1 and become progressively bigger at higher
temperatures. TQRPA comparison is similar as witnessed in upper left
panel. The mutual comparison between the five calculations remain
more or less same at stellar density of 10$^{9}$ gcm$^{-3}$ (lower
left panel). It is to be noted that at low temperatures the reported
electron capture rates are more than double the LSSM rates.  At
still higher density of 10$^{10}$ gcm$^{-3}$ (lower right panel) all
calculations agree with one another within a factor of ten as was in
the case of $^{54}$Fe (Fig.~\ref{figure4}). Figs.~\ref{figure4} and
\ref{figure6} depict the fact that the comparison between pn-QRPA
and LSSM is comparatively far better and that TQRPA model requires
further refinement. However during oxygen and silicon burning
phases,  the reported rates are a factor of 2-4 bigger than LSSM
calculated electron capture rates. Core-collapse simulators should
take note of these findings which can greatly assist in evolution of
fine-tuning of time rate of change of lepton fraction in their
codes.

\begin{figure}[htbp]
\begin{center}
\includegraphics[width=0.6\textwidth]{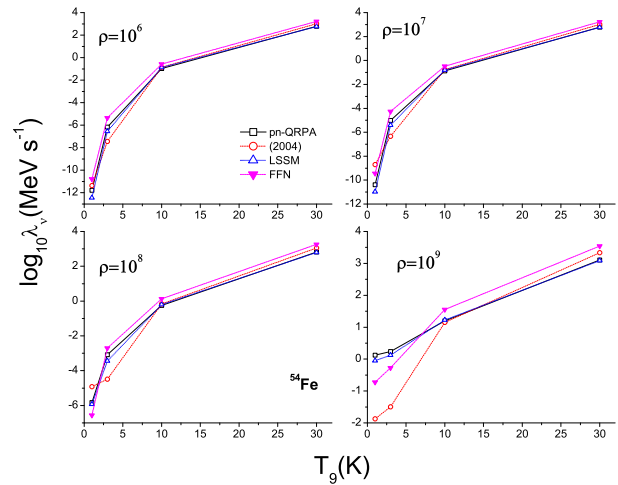}
\caption{\scriptsize Comparison of neutrino cooling rates due to
$^{54}$Fe using current pn-QRPA model (pn-QRPA), previous pn-QRPA
calculation (2004) \cite{Nab04}, shell model calculation (LSSM)
\cite{Lan00} and calculation done by Ref. \cite{Ful82} (FFN) at
different selected stellar temperature and density
zones.}\label{figure7}
\end{center}
\end{figure}

\begin{figure}[htbp]
\begin{center}
\includegraphics[width=0.6\textwidth]{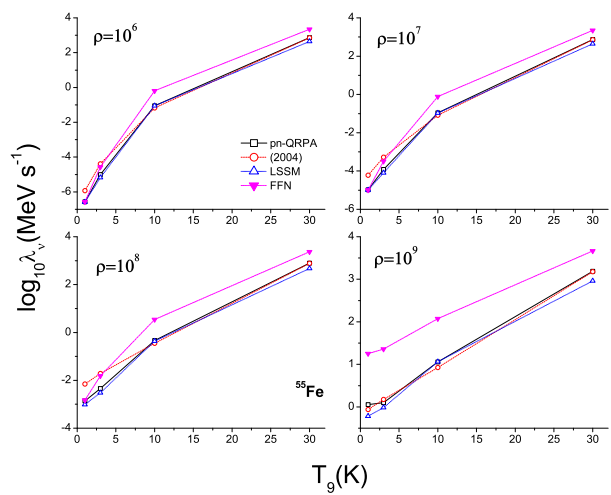}
\caption{\scriptsize Same as Figure~\ref{figure7} but for
$^{55}$Fe.} \label{figure8}
\end{center}
\end{figure}

\begin{figure}[htbp]
\begin{center}
\includegraphics[width=0.6\textwidth]{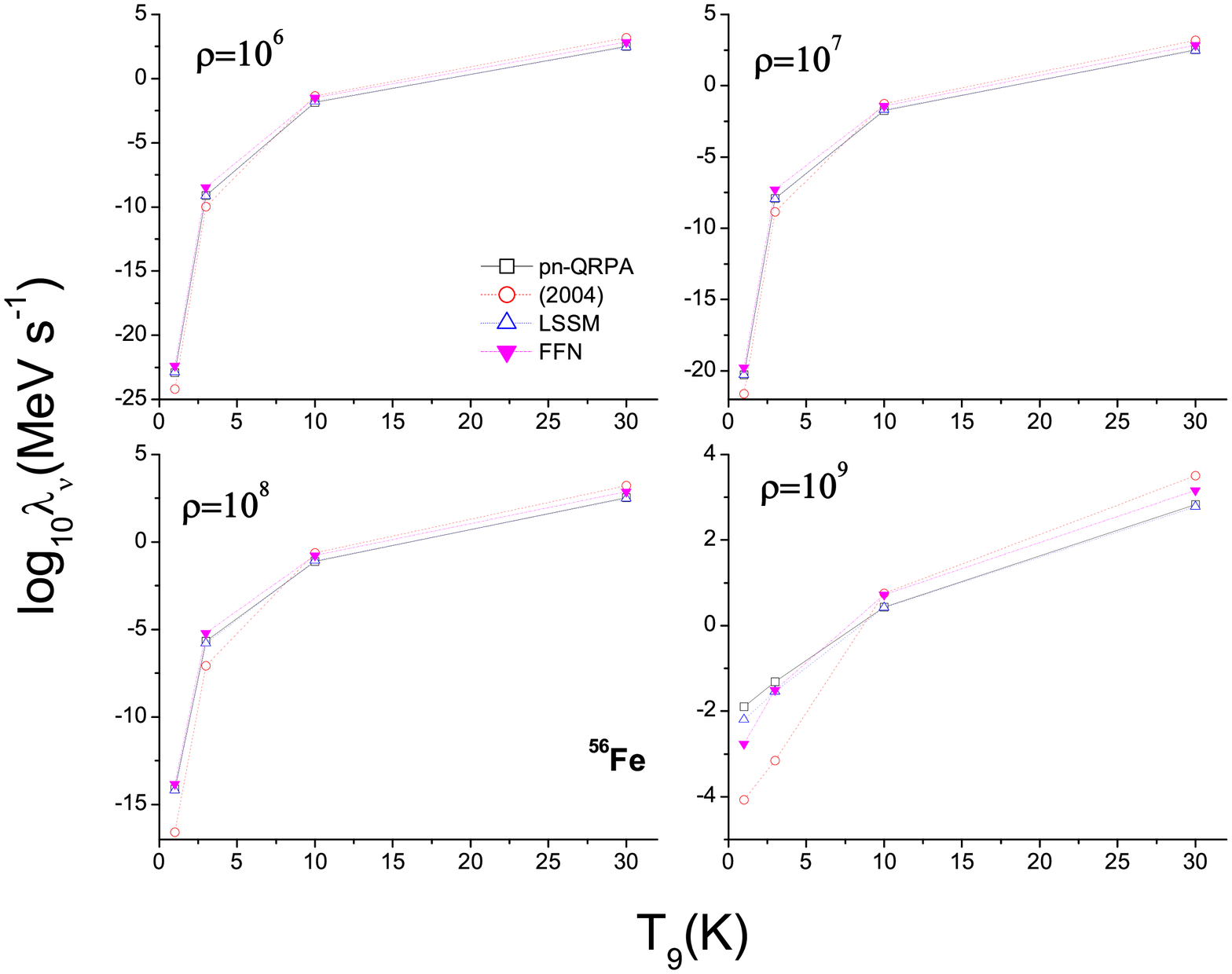}
\caption{\scriptsize Same as Figure~\ref{figure7} but for
$^{56}$Fe.}\label{figure9}
\end{center}
\end{figure}

Neutrino cooling rates play a key role in controlling the overall
entropy of the stellar matter.  As mentioned earlier the calculated
neutrino cooling rates contain contributions \textit{both} due to
electron capture \textit{and} positron decay on iron isotopes.
Whereas electron capture rates are orders of magnitude bigger than
competing positron decay rates, nonetheless they have a finite
contribution to the total neutrino cooling rates and as such are
included in the calculation. Figs.~\ref{figure7}-\ref{figure9} show
four panels depicting neutrino cooling rates due to electron capture
and positron decay rates on $^{54-56}$Fe. The upper left, upper
right, lower left and lower right panels show neutrino cooling rates
calculated at stellar densities of 10$^{6}$, 10$^{7}$, 10$^{8}$ and
10$^{9}$ gcm$^{-3}$, respectively. All neutrino cooling rates are
shown in logarithmic scale (to base 10) and calculated in units of
$MeV.s^{-1}$. Again the density scale in four panels were carefully
selected  where calculated electron capture \textit{and} positron
decay rates are most effective due to isotopes of iron as per the
modeling and simulation results of presupernova evolution of massive
stars by Heger and collaborators \cite{Heg01}. Consequently, for the
chosen temperature and density zone, the neutrino cooling rates
should be most effective in stellar matter.

Fig.~\ref{figure7} show reported neutrino cooling rates due to
$^{54}$Fe in stellar matter. Shown also are the previous pn-QRPA
cooling rates (2004) as well as LSSM and FFN results. It is
pertinent to mention that TQRPA formalism was not used to calculate
the neutrino cooling rates. Rates in upper left panel (plotted at a
stellar density of around 10$^{6}$ gcm$^{-3}$) show that the
previous pn-QRPA rates are, in general, bigger than reported rates
by an order of magnitude (except at T$_{9}$ =3 where the reported
rates are bigger). At T$_{9}$ =1 (3) the reported neutrino cooling
rates is around four (three) times the LSSM cooling rates. The
comparison with LSSM is perfect at higher temperatures. A more or
less similar comparison between the different theoretical
calculations exist at a stellar density of around 10$^{7}$
gcm$^{-3}$ (upper right panel). For still higher density of around
10$^{8}$ gcm$^{-3}$ (lower left panel) the reported rates are within
an order of magnitude difference with previous pn-QRPA rates. The
reported rates are around five times bigger than FFN rates at
T$_{9}$ =1 and smaller at all higher temperatures (by an order of
magnitude). At T$_{9}$ =3, the reported rates are twice the LSSM
rates. Otherwise the reported rates are in complete agreement with
LSSM reported neutrino cooling rates. The reported rates are 1-2
orders of magnitude bigger than previous pn-QRPA rates at low
temperatures (T$_{9} \le $ 3) and density of 10$^{9}$ gcm$^{-3}$
(lower right panel). At high temperatures the previous pn-QRPA rates
surpass the reported rates. Similarly the reported rates are a
factor 3-7 bigger at low temperatures (T$_{9} \le $ 3) compared to
FFN rates. At higher temperatures the FFN rates become bigger. LSSM
calculated neutrino cooling rates due to $^{54}$Fe are in perfect
agreement with reported rates at a stellar density of 10$^{9}$
gcm$^{-3}$. Overall the new pn-QRPA model results are more reliable
than previous pn-QRPA calculation (as also reflected in the good
comparison of the calculated GT strength distribution with
measurements shown earlier) and are attributed to a judicious choice
of model parameters stated earlier. FFN cooling rates are generally
bigger by an order of magnitude difference. The reason for this
artificial enhancement in FFN rates were also discussed earlier. The
new pn-QRPA model and LSSM calculations are in very good agreement,
both being microscopic in nature.

The calculated neutrino cooling rates due to $^{55}$Fe using
different nuclear models are shown in Fig.~\ref{figure8}. The
results are shown in a four panel system akin to Fig.~\ref{figure7}.
The reported neutrino cooling rates is roughly within a factor of
ten as compared to previous pn-QRPA and FFN calculations and are
found to be in excellent agreement with LSSM calculation. The origin
of these differences are already being discussed above.

We finally present the calculations of neutrino cooling rates due to
$^{56}$Fe in Fig.~\ref{figure9}. For this case the reported cooling
rates are 1-2 orders of magnitude bigger than previous pn-QRPA
calculation at low temperatures (T$_{9} \le $ 3). The comparison
improves at higher temperatures. FFN calculations are within a
factor 10 of reported rates while LSSM rates are once again in very
good agreement with reported rates for reasons already discussed.
During silicon shell burning stages and in later phases the reported
neutrino cooling rates is roughly double the LSSM rates. This
information may also prove useful for core-collapse simulators.
Generally speaking the new pn-QRPA calculated weak rates are in
excellent agreement with LSSM rates but at times under some very
specific physical conditions (already mentioned) there are some
minor differences. Incidentally it is precisely the fine-tuning of
the time evolution of lepton-to-baryon fraction (controlled by the
stellar weak interaction rates) which may be the key to a successful
explosion in the modeling and simulation of core-collapse
supernovae.
\section{Conclusions}
We presented a microscopic calculation of GT strength distribution
functions as well as electron capture and neutrino cooling rates due
to key iron isotopes in stellar matter. The calculations were also
compared  with experimental data and other theoretical calculations.
These nuclear inputs should prove useful for core-collapse
simulators.

The pn-QRPA model places the centroid of the calculated GT strength
distributions at much lower energies in daughter nuclei as compared
to other RPA interactions. Further the placement of GT centroids by
the pn-QRPA model is, in general, in good agreement with the
centroids of the measured data. This tendency of pn-QRPA model favor
higher values of electron capture rates in stellar environment and
can bear significance for astrophysical problems. On the other
extreme, the GXPF1 interaction usually leads to placement of GT
centroid at much higher energies in daughter compared to other
pn-RPA interactions. The reported total GT strengths are, in
general, also in good comparison with measured strengths.

During the oxygen and silicon shell and core burning phases of
massive stars the pn-QRPA electron capture (neutrino cooling) rates
on $^{54}$Fe are around three times (up to four times)  bigger than
those calculated by LSSM. The comparison with LSSM gets better for
proceeding pre-supernova and supernova phases of stars. The pn-QRPA
calculated electron capture rates on $^{55,56}$Fe are in overall
excellent agreement with the LSSM rates.  During silicon shell
burning for stars ($\sim 10 - 25M_{\odot}$) and oxygen shell burning
for much heavier stars ($\sim 40M_{\odot}$) the calculated cooling
rates due to $^{55}$Fe are in very good comparison with the LSSM
rates. The results for neutrino cooling rates due to $^{56}$Fe are
in overall good agreement with the corresponding LSSM numbers. FFN
rates, in general, are 1-2 orders of magnitude bigger.

The weak rates were calculated on a fine-grid scale, of great
utility for core-collapse simulators, and also suitable for
interpolation purposes. We present a sample of our fine-grid scale
calculation of weak rates at a stellar density of  10$^{11}$
gcm$^{-3}$ in Table~\ref{ta5}. The ASCII file of the complete table
can be requested from the corresponding author. It is expected that
the present calculation can contribute to a better understanding of
the explosion mechanism of massive stars. However significant
progress is expected, in part, to come from next-generation
radioactive ion-beam facilities (e.g. FAIR (Germany), FRIB (USA) and
FRIB (Japan)) when we would have access to measured GT strength
distribution of many more nuclei (including unstable isotopes).

\ack J.-U. Nabi would like to acknowledge the support provided by
TWAS and the local hospitality provided by the Egyptian Center for
Theoretical Physics, Cairo, where this project was partially
completed.

\newpage
\begin{table}
\caption{\scriptsize Calculated GT strength distributions in the
electron capture direction for $^{54}$Fe using different pn-RPA
scenarios. For details of interactions used see text. Calculated GT
strength less than $10^{-3}$ is not given to save space. Measured
data (Exp) is taken from Ref. \cite{Roe93}.} \label{ta1}
\begin{center}
\scriptsize \begin{tabular}{cc|cc|cc|cc|cc} \hline\hline
\multicolumn{2}{c|}{KB3G}& \multicolumn{2}{|c|}{GXPF1} &
\multicolumn{2}{|c|}{FPD6} & \multicolumn{2}{|c|}{pn-QRPA} & \multicolumn{2}{c}{Exp}\\
\multicolumn{1}{c}{$E_{i}$} & \multicolumn{1}{c|}{$B(GT_{+})$} &
\multicolumn{1}{c}{$E_{i}$} & \multicolumn{1}{c|}{$B(GT_{+})$} &
\multicolumn{1}{c}{$E_{i}$} & \multicolumn{1}{c|}{$B(GT_{+})$} &
\multicolumn{1}{c}{$E_{i}$} & \multicolumn{1}{c|}{$B(GT_{+})$} &
\multicolumn{1}{c}{$E_{i}$} & \multicolumn{1}{c}{$B(GT_{+})$} \\
\hline

4.93 & 1.50 $\times 10^{-2}$  & 4.68 & 3.39 $\times 10^{-3}$ &
5.19 & 1.49 $\times 10^{-2}$ & 1.39 & 3.56 $\times 10^{-1}$ & -1.5 & 0.05 \\

4.93 & 1.18 $\times 10^{-2}$ & 4.68 & 3.39 $\times 10^{-3}$ &
5.19 & 1.49 $\times 10^{-2}$ & 1.45 & 6.33 $\times 10^{-2}$ & -0.5 & 0.12 \\

4.93 & 1.18 $\times 10^{-2}$ & 4.83 & 8.78 $\times 10^{-3}$ &
5.29 & 2.46 $\times 10^{-2}$ & 1.61 & 5.37 $\times 10^{-2}$ & 0.5 & 0.25 \\

5.41 & 4.32 $\times 10^{-3}$ & 5.05 & 7.52 $\times 10^{-3}$ &
5.54 & 5.37 $\times 10^{-2}$ & 1.78 & 3.01 $\times 10^{-2}$ & 1.5 & 0.45 \\

5.41 & 4.32 $\times 10^{-3}$ & 5.05 & 7.52 $\times 10^{-3}$ &
5.54 & 5.37 $\times 10^{-2}$ & 1.86 & 2.60 $\times 10^{-2}$ & 2.5 & 0.57 \\

5.60 & 7.30 $\times 10^{-2}$ & 5.67 & 1.80 $\times 10^{-2}$ &
5.73 & 3.04 $\times 10^{-2}$ & 2.24 & 1.90 $\times 10^{-2}$ & 3.5 & 0.54 \\

5.60 & 7.30 $\times 10^{-2}$ & 5.67 & 1.80 $\times 10^{-2}$ &
5.73 & 3.04 $\times 10^{-2}$ & 2.36 & 1.08 $\times 10^{-1}$ & 4.5 & 0.51 \\

5.84 & 8.09 $\times 10^{-2}$ & 5.76 & 3.85 $\times 10^{-3}$ &
6.00 & 7.94 $\times 10^{-2}$  & 2.59 & 2.89 $\times 10^{-3}$ & 5.5 & 0.38 \\

5.84 & 8.09 $\times 10^{-2}$ & 5.82 & 2.28 $\times 10^{-2}$ &
6.00 & 7.94 $\times 10^{-2}$ & 2.70 & 2.83 $\times 10^{-2}$ & 6.5 & 0.41 \\

6.16 & 1.49 $\times 10^{-1}$ & 5.82 & 2.28 $\times 10^{-2}$ &
6.44 & 2.80 $\times 10^{-2}$ & 2.72 & 1.77 $\times 10^{-1}$ & 7.5 & 0.19 \\

6.18 & 8.04 $\times 10^{-2}$ & 6.05 & 2.22 $\times 10^{-2}$ &
6.44 & 2.80 $\times 10^{-2}$ & 2.97 & 5.43 $\times 10^{-2}$ & 8.5 & 0.05 \\

6.18 & 8.04 $\times 10^{-2}$ & 6.05 & 2.22 $\times 10^{-2}$ &
6.51 & 1.50 $\times 10^{-1}$ & 3.05 & 4.39 $\times 10^{-2}$ && \\

6.50 & 1.35 $\times 10^{-1}$ & 6.29 & 5.50 $\times 10^{-2}$ &
6.96 & 3.98 $\times 10^{-1}$ & 3.09 & 4.28 $\times 10^{-3}$ & & \\

6.50 & 1.35 $\times 10^{-1}$ & 7.11 & 3.99 $\times 10^{-2}$ &
6.96 & 3.98 $\times 10^{-1}$ & 3.40 & 2.70 $\times 10^{-2}$ & & \\

6.78 & 7.14 $\times 10^{-1}$ & 7.11 & 3.99 $\times 10^{-2}$ &
7.24 & 6.19 $\times 10^{-1}$ & 3.49 & 3.07 $\times 10^{-2}$ & & \\

6.78 & 7.14 $\times 10^{-1}$ & 7.70 & 5.12 $\times 10^{-1}$ &
7.24 & 6.19 $\times 10^{-1}$ & 3.57 & 1.93 $\times 10^{-1}$ & & \\

6.84 & 5.17 $\times 10^{-1}$ & 7.70 & 5.12 $\times 10^{-1}$  &
7.24 & 3.23 $\times 10^{-1}$ & 3.83 & 2.62 $\times 10^{-1}$ & & \\

7.32 & 2.73 $\times 10^{-2}$ & 7.72 & 4.25 $\times 10^{-1}$ &
8.13 & 3.73 $\times 10^{-2}$ & 3.89 & 3.53 $\times 10^{-2}$ & & \\

7.46 & 1.89 $\times 10^{-1}$ & 8.07 & 7.28 $\times 10^{-2}$&
8.13 & 3.73 $\times 10^{-2}$ & 3.90 & 1.16 $\times 10^{-3}$ & &\\

7.46 & 1.89 $\times 10^{-1}$ & 8.13 & 1.51 $\times 10^{-1}$ &
8.15 & 2.31 $\times 10^{-1}$ & 4.08 & 1.19 $\times 10^{-3}$ & & \\

7.67 & 3.86 $\times 10^{-2}$ & 8.13 & 1.51 $\times 10^{-1}$ &
8.44 & 3.08 $\times 10^{-2}$  & 4.15 & 4.05 $\times 10^{-2}$ & & \\

7.67 & 3.86 $\times 10^{-2}$ & 8.16 & 6.46 $\times 10^{-1}$ &
8.44 & 3.08 $\times 10^{-2}$ & 4.34 & 2.18 $\times 10^{-2}$ & & \\

8.16 & 9.56 $\times 10^{-2}$ &  8.16 & 6.46 $\times 10^{-1}$ &
8.80 & 3.31 $\times 10^{-2}$ & 4.38 & 2.22 $\times 10^{-3}$ & & \\

8.16 & 9.56 $\times 10^{-2}$ & 9.13 & 1.68 $\times 10^{-1}$   &
8.80 & 3.31 $\times 10^{-2}$ & 4.60 & 1.67 $\times 10^{-2}$ & & \\

8.30 & 1.07 $\times 10^{0}$ & 9.13 & 1.68 $\times 10^{-1}$ &
9.11 & 8.56 $\times 10^{-1}$ & 4.68 & 1.27 $\times 10^{-1}$ & & \\

& & 9.73 & 1.44 $\times 10^{0}$ & &
 & 4.92 & 7.14 $\times 10^{-3}$ & & \\

& &  & & &
 & 4.97 & 3.01 $\times 10^{-2}$ & & \\

& &  & & &
 & 5.03 & 1.84 $\times 10^{-3}$ & & \\

& & &  & &  & 5.25 & 5.38 $\times 10^{-2}$ & & \\
& & &  & &  & 5.38 & 1.50 $\times 10^{-1}$ & & \\
& & &  & &  & 5.61 & 5.68 $\times 10^{-3}$ & & \\
& & &  & &  & 5.64 & 1.11 $\times 10^{-3}$ & & \\
& & &  & &  & 5.71 & 3.44 $\times 10^{-2}$ & & \\
& & &  & &  & 5.77 & 2.26 $\times 10^{-2}$ & & \\
& & &  & &  & 5.83 & 8.02 $\times 10^{-3}$ & & \\
& & &  & &  & 5.92 & 5.56 $\times 10^{-2}$ & & \\
& & &  & &  & 6.12 & 4.14 $\times 10^{-3}$ & & \\
& & &  & &  & 6.14 & 2.66 $\times 10^{-2}$ & & \\
& & &  & &  & 6.15 & 1.05 $\times 10^{-2}$ & & \\
& & &  & &  & 6.24 & 1.09 $\times 10^{-2}$ & & \\
& & &  & &  & 6.44 & 1.22 $\times 10^{-3}$ & & \\
& & &  & &  & 6.47 & 2.65 $\times 10^{-2}$ & & \\
& & &  & &  & 6.50 & 5.61 $\times 10^{-2}$ & & \\
& & &  & &  & 6.86 & 1.15 $\times 10^{-2}$ & & \\
& & &  & &  & 6.97 & 3.52 $\times 10^{-3}$ & & \\
& & &  & &  & 7.13 & 4.45 $\times 10^{-1}$ & & \\
& & &  & &  & 7.56 & 1.25 $\times 10^{-3}$ & & \\
& & &  & &  & 7.80 & 1.23 $\times 10^{-3}$ & & \\
& & &  & &  & 7.85 & 1.74 $\times 10^{-3}$ & & \\
& & &  & &  & 10.17 & 4.76$\times 10^{-3}$ & & \\
& & &  & &  & 10.31 & 2.96$\times 10^{-2}$ & & \\
& & &  & &  & 10.49 & 9.95$\times 10^{-2}$ & & \\
& & &  & &  & 10.57 & 2.02$\times 10^{-3}$ & & \\
& & &  & &  & 10.88 & 4.90$\times 10^{-1}$ & & \\
& & &  & &  & 10.98 & 5.92$\times 10^{-3}$ & & \\
& & &  & &  & 11.06 & 2.91$\times 10^{-2}$ & & \\
& & &  & &  & 11.12 & 1.89$\times 10^{-1}$ & & \\
& & &  & &  & 11.26 & 3.90$\times 10^{-2}$ & & \\
\end{tabular}
\end{center}
\end{table}

\begin{table}
\caption{\scriptsize Same as Table~\ref{ta1} but for $^{55}$Fe.}
\label{ta2}
\begin{center}
\scriptsize \begin{tabular}{cc|cc|cc|cc} \hline\hline
\multicolumn{2}{c|}{KB3G}& \multicolumn{2}{|c|}{GXPF1} &
\multicolumn{2}{|c|}{FPD6} & \multicolumn{2}{|c}{pn-QRPA} \\
\multicolumn{1}{c}{$E_{i}$} & \multicolumn{1}{c|}{$B(GT_{+})$} &
\multicolumn{1}{c}{$E_{i}$} & \multicolumn{1}{c|}{$B(GT_{+})$} &
\multicolumn{1}{c}{$E_{i}$} & \multicolumn{1}{c|}{$B(GT_{+})$} &
\multicolumn{1}{c}{$E_{i}$} & \multicolumn{1}{c}{$B(GT_{+})$} \\
\hline

0.65 & 2.75 $\times 10^{-3}$ & 0.21 & 5.28 $\times 10^{-3}$ &
0.52 & 7.30 $\times 10^{-3}$ & 0.00 & 2.55 $\times 10^{-3}$  \\

0.66 & 3.75 $\times 10^{-3}$ & 0.23 & 8.35 $\times 10^{-3}$ &
0.58 & 1.14 $\times 10^{-2}$ & 1.29 & 2.88 $\times 10^{-2}$  \\

1.27 & 4.70 $\times 10^{-3}$ & 0.67 & 1.01 $\times 10^{-2}$ &
1.03 & 5.80 $\times 10^{-3}$ & 2.17 & 1.13 $\times 10^{-1}$  \\

1.30 & 1.05 $\times 10^{-2}$ & 1.00 & 3.30 $\times 10^{-2}$ &
1.05 & 1.49 $\times 10^{-2}$ & 2.57 & 2.76 $\times 10^{-2}$  \\

1.58 & 4.45 $\times 10^{-2}$ & 1.41 & 4.42 $\times 10^{-3}$ &
1.30 & 8.94 $\times 10^{-3}$ & 2.66 & 1.91 $\times 10^{-3}$  \\

1.70 & 1.51 $\times 10^{-2}$ & 1.41 & 4.20 $\times 10^{-3}$ &
1.79 & 3.52 $\times 10^{-2}$ & 2.94 & 3.71 $\times 10^{-2}$  \\

1.70 & 1.59 $\times 10^{-2}$ & 1.60 & 1.32 $\times 10^{-2}$ &
1.82 & 3.20 $\times 10^{-2}$ & 3.88 & 4.40 $\times 10^{-3}$  \\

2.07 & 1.86 $\times 10^{-2}$ & 1.60 & 1.36 $\times 10^{-2}$ &
1.93 & 5.47 $\times 10^{-2}$ & 4.70 & 2.08 $\times 10^{-3}$  \\

2.07 & 2.51 $\times 10^{-2}$ & 2.10 & 6.48 $\times 10^{-3}$ &
2.24 & 8.67 $\times 10^{-2}$ & 4.98 & 4.27 $\times 10^{-2}$  \\

2.19 & 9.45 $\times 10^{-3}$ & 2.10 & 1.56 $\times 10^{-2}$ &
2.25 & 9.90 $\times 10^{-2}$ & 5.05 & 7.64 $\times 10^{-2}$  \\

2.22 & 9.18 $\times 10^{-2}$ & 2.20 & 2.27 $\times 10^{-2}$ &
2.41 & 2.12 $\times 10^{-3}$ & 5.06 & 3.81 $\times 10^{-3}$  \\

2.29 & 1.67 $\times 10^{-1}$ & 2.35 & 5.43 $\times 10^{-3}$ &
2.43 & 1.40 $\times 10^{-2}$ & 5.22 & 1.47 $\times 10^{-1}$ \\

2.32 & 1.98 $\times 10^{-2}$ & 2.37 & 5.47 $\times 10^{-3}$ &
2.69 & 4.54 $\times 10^{-2}$ & 5.28 & 1.83 $\times 10^{-2}$  \\

2.52 & 8.75 $\times 10^{-3}$ & 2.39 & 2.56 $\times 10^{-3}$ &
2.70 & 1.09 $\times 10^{-1}$ & 5.32 & 1.81 $\times 10^{-3}$  \\

2.62 & 3.19 $\times 10^{-2}$ & 2.45 & 3.42 $\times 10^{-2}$ &
2.89 & 3.32 $\times 10^{-2}$ & 5.41 & 8.02 $\times 10^{-3}$  \\

2.79 & 1.62 $\times 10^{-3}$ & 2.53 & 3.85 $\times 10^{-2}$ &
2.92 & 7.46 $\times 10^{-3}$ & 5.57 & 6.72 $\times 10^{-2}$  \\

2.79 & 9.10 $\times 10^{-3}$ & 2.72 & 1.17 $\times 10^{-3}$ &
3.06 & 1.38 $\times 10^{-3}$ & 6.68 & 1.39 $\times 10^{-1}$  \\

2.80 & 2.76 $\times 10^{-3}$ & 2.86 & 6.17 $\times 10^{-3}$ &
3.09 & 5.68 $\times 10^{-3}$ & 6.76 & 3.80 $\times 10^{-2}$  \\

2.85 & 2.01 $\times 10^{-2}$ & 2.88 & 4.12 $\times 10^{-3}$ &
3.16 & 3.72 $\times 10^{-2}$ & 6.87 & 1.90 $\times 10^{-1}$ \\

2.92 & 4.36 $\times 10^{-2}$ & 2.89 & 5.24 $\times 10^{-3}$ &
3.21 & 2.33 $\times 10^{-3}$ & 6.89 & 5.94 $\times 10^{-2}$  \\

2.99 & 3.16 $\times 10^{-2}$ & 3.04 & 2.24 $\times 10^{-2}$ &
3.36 & 1.01 $\times 10^{-3}$ & 7.04 & 2.83 $\times 10^{-2}$  \\

3.02 & 1.97 $\times 10^{-2}$ & 3.12 & 4.62 $\times 10^{-3}$ &
3.43 & 2.26 $\times 10^{-1}$ & 7.30 & 3.19 $\times 10^{-1}$ \\

3.17 & 1.40 $\times 10^{-2}$ & 3.12 & 4.75 $\times 10^{-3}$ &
3.53 & 5.98 $\times 10^{-2}$ & 7.35 & 8.70 $\times 10^{-3}$  \\

3.23 & 5.88 $\times 10^{-2}$ & 3.30 & 6.75 $\times 10^{-2}$ &
3.80 & 3.52 $\times 10^{-2}$ & 7.54 & 4.33 $\times 10^{-2}$  \\

3.33 & 6.74 $\times 10^{-2}$ & 3.42 & 2.14 $\times 10^{-2}$ &
3.98 & 1.51 $\times 10^{-1}$ & 7.62 & 1.13 $\times 10^{-2}$  \\

3.39 & 1.40 $\times 10^{-1}$ & 3.45 & 2.43 $\times 10^{-2}$ &
4.11 & 3.50 $\times 10^{-1}$ & 7.66 & 1.60 $\times 10^{-2}$  \\

3.50 & 1.75 $\times 10^{-1}$ & 3.76 & 8.23 $\times 10^{-2}$ &
4.29 & 1.40 $\times 10^{-1}$ & 7.72 & 9.35 $\times 10^{-2}$  \\

3.66 & 4.00 $\times 10^{-1}$ & 3.79 & 7.81 $\times 10^{-2}$ &
4.36 & 8.62 $\times 10^{-2}$ & 7.75 & 6.00 $\times 10^{-2}$  \\

3.73 & 2.32 $\times 10^{-1}$ & 3.96 & 3.30 $\times 10^{-2}$ &
4.39 & 1.48 $\times 10^{-1}$ & 8.00 & 5.78 $\times 10^{-1}$  \\

3.77 & 6.56 $\times 10^{-2}$ & 4.03 & 4.13 $\times 10^{-2}$ &
4.45 & 1.85 $\times 10^{-2}$ & 8.02 & 3.44 $\times 10^{-2}$  \\

3.77 & 1.59 $\times 10^{-1}$ & 4.09 & 1.10 $\times 10^{-2}$ &
4.46 & 2.31 $\times 10^{-2}$ & 8.07 & 4.51 $\times 10^{-3}$  \\

3.81 & 1.12 $\times 10^{-1}$ & 4.15 & 2.97 $\times 10^{-1}$ &
4.61 & 1.25 $\times 10^{-1}$ & 8.10 & 6.44 $\times 10^{-2}$  \\

3.92 & 1.34 $\times 10^{-1}$ & 4.19 & 7.81 $\times 10^{-2}$ &
4.70 & 8.06 $\times 10^{-2}$ & 8.21 & 5.66 $\times 10^{-3}$  \\

3.95 & 1.54 $\times 10^{-1}$ & 4.28 & 8.11 $\times 10^{-2}$ &
4.78 & 6.59 $\times 10^{-2}$ & 8.27 & 1.65 $\times 10^{-2}$  \\

4.00 & 3.74 $\times 10^{-3}$ & 4.30 & 1.52 $\times 10^{-2}$ &
4.81 & 1.51 $\times 10^{-2}$ & 8.28 & 1.43 $\times 10^{-2}$  \\

4.17 & 3.87 $\times 10^{-2}$ & 4.31 & 3.61 $\times 10^{-2}$ &
4.93 & 1.63 $\times 10^{-2}$ & 8.30 & 4.52 $\times 10^{-3}$  \\

4.19 & 6.99 $\times 10^{-2}$ & 4.43 & 2.24 $\times 10^{-1}$ &
5.13 & 2.87 $\times 10^{-2}$ & 8.36 & 2.52 $\times 10^{-1}$  \\

4.23 & 1.16 $\times 10^{-3}$ & 4.47 & 1.32 $\times 10^{-1}$ &
5.13 & 8.34 $\times 10^{-2}$ & 8.36 & 4.62 $\times 10^{-2}$  \\

4.23 & 1.99 $\times 10^{-2}$ & 4.72 & 3.63 $\times 10^{-1}$ &
5.19 & 6.48 $\times 10^{-2}$ & 8.43 & 1.14 $\times 10^{-1}$  \\

4.40 & 8.37 $\times 10^{-2}$ & 4.81 & 6.58 $\times 10^{-2}$ &
5.20 & 3.24 $\times 10^{-2}$ & 8.65 & 1.68 $\times 10^{-3}$  \\

4.41 & 6.65 $\times 10^{-2}$ & 4.85 & 5.28 $\times 10^{-3}$ &
5.59 & 2.53 $\times 10^{-3}$ & 8.85 & 6.87 $\times 10^{-3}$  \\

4.59 & 1.95 $\times 10^{-2}$ & 4.92 & 3.33 $\times 10^{-1}$ &
5.60 & 5.20 $\times 10^{-2}$ & 8.93 & 6.01 $\times 10^{-2}$  \\

4.59 & 1.36 $\times 10^{-2}$ & 4.94 & 5.36 $\times 10^{-2}$ &
5.66 & 3.06 $\times 10^{-3}$ & 8.95 & 1.28 $\times 10^{-1}$  \\

4.89 & 1.42 $\times 10^{-2}$ & 5.04 & 2.19 $\times 10^{-1}$ &
5.90 & 2.90 $\times 10^{-1}$ & 9.17 & 2.60 $\times 10^{-3}$  \\

4.89 & 2.37 $\times 10^{-1}$ & 5.21 & 3.10 $\times 10^{-1}$ &
6.01 & 3.06 $\times 10^{-3}$ & 9.25 & 3.18 $\times 10^{-3}$  \\

4.98 & 2.48 $\times 10^{-2}$ & 5.24 & 1.83 $\times 10^{-1}$ &
6.35 & 2.82 $\times 10^{-1}$ & 9.30 & 4.77 $\times 10^{-3}$  \\

5.09 & 2.67 $\times 10^{-2}$ & 5.46 & 5.88 $\times 10^{-2}$ &
6.39 & 3.54 $\times 10^{-2}$ & 9.51 & 1.43 $\times 10^{-2}$  \\

5.12 & 2.66 $\times 10^{-1}$ & 5.59 & 9.05 $\times 10^{-3}$ &
6.47 & 3.22 $\times 10^{-2}$ & 9.66 & 1.85 $\times 10^{-3}$  \\

5.21 & 5.22 $\times 10^{-2}$ & 5.63 & 8.05 $\times 10^{-2}$ &
6.56 & 1.67 $\times 10^{-2}$ & 9.69 & 1.08 $\times 10^{-3}$  \\

5.53 & 9.35 $\times 10^{-3}$ & 5.95 & 1.32 $\times 10^{-2}$ &
6.78 & 7.69 $\times 10^{-3}$ & 9.69 & 5.38 $\times 10^{-3}$  \\

5.54 & 5.59 $\times 10^{-2}$ & 6.09 & 2.54 $\times 10^{-3}$ &
7.03 & 1.19 $\times 10^{-2}$ & 10.05 & 6.35 $\times 10^{-3}$  \\

5.60 & 3.20 $\times 10^{-1}$ & 6.19 & 1.35 $\times 10^{-1}$ &
7.06 & 7.30 $\times 10^{-2}$ & 10.06 & 2.60 $\times 10^{-3}$  \\

5.63 & 2.76 $\times 10^{-2}$ & 6.24 & 3.05 $\times 10^{-2}$ &
7.27 & 1.71 $\times 10^{-3}$ & 10.09 & 7.04 $\times 10^{-3}$  \\

5.70 & 1.84 $\times 10^{-2}$ & 6.40 & 1.89 $\times 10^{-1}$ &
7.30 & 3.45 $\times 10^{-3}$ & 10.16 & 1.81 $\times 10^{-3}$  \\

5.74 & 2.73 $\times 10^{-2}$ & 6.46 & 2.78 $\times 10^{-2}$ &
7.64 & 2.48 $\times 10^{-2}$ & 13.97 & 1.37 $\times 10^{-3}$  \\

5.96 & 7.45 $\times 10^{-2}$ & 6.65 & 7.02 $\times 10^{-1}$ &
7.90 & 2.08 $\times 10^{-3}$ & &  \\

6.06 & 2.99 $\times 10^{-3}$ & 6.81 & 1.27 $\times 10^{-2}$ &
& & &  \\

6.16 & 3.06 $\times 10^{-2}$ & 6.84 & 2.22 $\times 10^{-1}$ &
& & &  \\

6.23 & 2.21 $\times 10^{-3}$ & 7.19 & 2.46 $\times 10^{-3}$  &
& & &  \\

6.49 & 1.88 $\times 10^{-3}$ & 7.19 & 1.05 $\times 10^{-2}$ &
& & &  \\
\end{tabular}
\end{center}
\end{table}

\begin{table}
\caption{\scriptsize Same as Table~\ref{ta1} but for $^{56}$Fe.
Measured data (Exp) is taken from Ref. \cite{Elk94}.} \label{ta3}
\begin{center}
\scriptsize \begin{tabular}{cc|cc|cc|cc|cc} \hline\hline
\multicolumn{2}{c|}{KB3G}& \multicolumn{2}{|c|}{GXPF1} &
\multicolumn{2}{|c|}{FPD6} & \multicolumn{2}{|c|}{pn-QRPA} & \multicolumn{2}{c}{Exp}\\
\multicolumn{1}{c}{$E_{i}$} & \multicolumn{1}{c|}{$B(GT_{+})$} &
\multicolumn{1}{c}{$E_{i}$} & \multicolumn{1}{c|}{$B(GT_{+})$} &
\multicolumn{1}{c}{$E_{i}$} & \multicolumn{1}{c|}{$B(GT_{+})$} &
\multicolumn{1}{c}{$E_{i}$} & \multicolumn{1}{c|}{$B(GT_{+})$} &
\multicolumn{1}{c}{$E_{i}$} & \multicolumn{1}{c}{$B(GT_{+})$} \\
\hline

3.46 & 3.12 $\times 10^{-1}$  & 3.83 & 1.46 $\times 10^{-1}$ &
3.90 & 3.69 $\times 10^{-1}$ & 0.11 & 3.82 $\times 10^{-2}$ & -2.0 & 0.0084 \\

3.46 & 3.12 $\times 10^{-1}$ & 3.83 & 1.46 $\times 10^{-1}$ &
3.90 & 3.69 $\times 10^{-1}$ & 0.22 & 2.53 $\times 10^{-1}$ & -1.0 & 0.13 \\

4.80 & 2.75 $\times 10^{-2}$ & 4.94 & 2.13 $\times 10^{-2}$ &
5.27 & 8.48 $\times 10^{-3}$ & 0.77 & 2.42 $\times 10^{-3}$ & 0.0 & 0.40 \\

4.86 & 2.54 $\times 10^{-1}$ & 5.36 & 6.89 $\times 10^{-2}$ &
5.27 & 8.48 $\times 10^{-3}$ & 0.83 & 1.14 $\times 10^{-2}$ & 1.0 & 0.57 \\

4.86 & 2.54 $\times 10^{-1}$ & 5.36 & 6.89 $\times 10^{-2}$ &
5.83 & 9.83 $\times 10^{-2}$ & 1.18 & 9.14 $\times 10^{-3}$ & 2.0 & 0.56 \\

5.42 & 1.92 $\times 10^{-1}$ & 5.51 & 8.41 $\times 10^{-2}$ &
5.97 & 2.28 $\times 10^{-1}$ & 1.24 & 1.00 $\times 10^{-2}$ & 3.0 & 0.29 \\

5.42 & 2.40 $\times 10^{-1}$ & 5.95 & 1.57 $\times 10^{-1}$ &
5.97 & 2.28 $\times 10^{-1}$ & 1.29 & 4.44 $\times 10^{-2}$ & 4.0 & 0.27 \\

5.42 & 2.40 $\times 10^{-1}$ & 5.95 & 1.56 $\times 10^{-1}$ &
6.72 & 6.62 $\times 10^{-2}$ & 1.56 & 1.55 $\times 10^{-1}$ & 5.0 & 0.15 \\

5.88 & 9.99 $\times 10^{-2}$ & 6.02 & 4.88 $\times 10^{-1}$ &
6.76 & 1.32 $\times 10^{-1}$ & 1.73 & 5.53 $\times 10^{-3}$ & 6.0 & 0.14 \\

5.88 & 9.99 $\times 10^{-2}$ & 6.02 & 4.88 $\times 10^{-1}$ &
6.76 & 1.32 $\times 10^{-1}$ & 1.98 & 4.79 $\times 10^{-2}$ & 7.0 & 0.14 \\

5.96 & 3.22 $\times 10^{-2}$ & 6.64 & 3.04 $\times 10^{-1}$ &
7.04 & 3.32 $\times 10^{-2}$ & 2.31 & 2.52 $\times 10^{-2}$ & 8.0 & 0.19 \\

6.56 & 7.62 $\times 10^{-3}$ & 6.73 & 1.86 $\times 10^{-2}$ &
7.04 & 3.32 $\times 10^{-2}$ & 2.48 & 2.62 $\times 10^{-3}$ &  9.0 & 0.17 \\

6.56 & 7.62 $\times 10^{-3}$ & 6.73 & 1.86 $\times 10^{-2}$ &
7.19 & 2.89 $\times 10^{-2}$ & 2.55 & 5.90 $\times 10^{-2}$ &  10.0 & 0.22 \\

6.66 & 6.95 $\times 10^{-1}$ & 7.40 & 2.56 $\times 10^{-1}$ &
8.10 & 5.14 $\times 10^{-1}$ & 2.62 & 4.39 $\times 10^{-2}$ & & \\

7.21 & 6.04 $\times 10^{-2}$ & 7.40 & 2.56 $\times 10^{-1}$ &
8.77 & 2.47 $\times 10^{-2}$ & 2.67 & 7.38 $\times 10^{-2}$ & & \\

7.21 & 6.04 $\times 10^{-2}$ & 7.75 & 1.28 $\times 10^{-2}$ &
8.77 & 2.47 $\times 10^{-2}$ & 2.77 & 3.70 $\times 10^{-2}$ & & \\

7.71 & 1.02 $\times 10^{-2}$ & 7.97 & 7.79 $\times 10^{-3}$  &
9.18 & 1.58 $\times 10^{-2}$ & 2.77 & 6.00 $\times 10^{-2}$ & & \\

7.71 & 1.02 $\times 10^{-2}$ & 7.97 & 7.79 $\times 10^{-3}$ &
9.18 & 1.58 $\times 10^{-2}$ & 2.79 & 7.36 $\times 10^{-2}$ & & \\

7.90 & 1.51 $\times 10^{-1}$ & 8.48 & 9.14 $\times 10^{-1}$&
9.89 & 5.12 $\times 10^{-2}$ & 2.89 & 3.16 $\times 10^{-3}$ & &\\

8.09 & 1.90 $\times 10^{-2}$ & 8.84 & 4.14 $\times 10^{-2}$ &
10.10 & 1.69 $\times 10^{-3}$ & 3.00 & 2.11 $\times 10^{-2}$ & & \\

8.09 & 1.90 $\times 10^{-2}$ & 8.84 & 4.14 $\times 10^{-2}$ &
10.10 & 1.69 $\times 10^{-3}$  & 3.07 & 1.07 $\times 10^{-2}$ & & \\

& & &  & &  & 3.22 & 8.47 $\times 10^{-3}$ & & \\
& & &  & &  & 3.26 & 3.39 $\times 10^{-2}$ & & \\
& & &  & &  & 3.33 & 1.34 $\times 10^{-3}$ & & \\
& & &  & &  & 3.75 & 9.74 $\times 10^{-2}$ & & \\
& & &  & &  & 3.78 & 2.55 $\times 10^{-3}$ & & \\
& & &  & &  & 3.86 & 1.59 $\times 10^{-3}$ & & \\
& & &  & &  & 4.02 & 8.90 $\times 10^{-1}$ & & \\
& & &  & &  & 4.33 & 5.49 $\times 10^{-2}$ & & \\
& & &  & &  & 4.54 & 7.37 $\times 10^{-2}$ & & \\
& & &  & &  & 4.66 & 2.17 $\times 10^{-2}$ & & \\
& & &  & &  & 4.70 & 2.62 $\times 10^{-2}$ & & \\
& & &  & &  & 4.75 & 1.63 $\times 10^{-2}$ & & \\
& & &  & &  & 4.83 & 1.15 $\times 10^{-3}$ & & \\
& & &  & &  & 5.03 & 2.14 $\times 10^{-2}$ & & \\
& & &  & &  & 5.08 & 4.90 $\times 10^{-3}$ & & \\
& & &  & &  & 5.13 & 8.44 $\times 10^{-3}$ & & \\
& & &  & &  & 5.15 & 2.68 $\times 10^{-3}$ & & \\
& & &  & &  & 5.23 & 2.47 $\times 10^{-3}$ & & \\
& & &  & &  & 5.31 & 1.54 $\times 10^{-2}$ & & \\
& & &  & &  & 5.43 & 1.58 $\times 10^{-3}$ & & \\
& & &  & &  & 5.47 & 1.84 $\times 10^{-2}$ & & \\
& & &  & &  & 5.53 & 1.80 $\times 10^{-3}$ & & \\
& & &  & &  & 6.07 & 5.96 $\times 10^{-3}$ & & \\
& & &  & &  & 6.07 & 1.06 $\times 10^{-2}$ & & \\
& & &  & &  & 6.10 & 2.23 $\times 10^{-2}$ & & \\
& & &  & &  & 6.15 & 5.12 $\times 10^{-3}$ & & \\
& & &  & &  & 6.28 & 1.43 $\times 10^{-3}$ & & \\
& & &  & &  & 6.49 & 2.95 $\times 10^{-3}$ & & \\
& & &  & &  & 6.55 & 4.98 $\times 10^{-3}$ & & \\
& & &  & &  & 6.83 & 1.01 $\times 10^{-3}$ & & \\
& & &  & &  & 7.06 & 3.05 $\times 10^{-3}$ & & \\
\end{tabular}
\end{center}
\end{table}

\begin{table}
\caption{\scriptsize Statistics of calculated GT strength
distributions for nuclei using different RPA models given in second
column. For references of theoretical calculations see text, while
experimental references are: (a) $\rightarrow$ \cite{Roe93}, (b)
$\rightarrow$ \cite{Vet89}, (c) $\rightarrow$ \cite{Ada12}, (d)
$\rightarrow$ \cite{Rap83}, (e) $\rightarrow$ \cite{And90}, (f)
$\rightarrow$ \cite{Elk94}.} \label{ta4}
\begin{center}
\scriptsize \begin{tabular}{c|c|ccc|ccc} Nucleus &
Model & E(GT$_{+}$) & Width(GT$_{+}$) & $\sum B(GT_+)$  & E(GT$_{-}$) & Width(GT$_{-}$) &  $\sum B(GT_-)$ \\
& & MeV & MeV & arb. units & MeV & MeV & arb. units \\
\hline $^{54}$Fe & & & & & & &\\
& FPD6  & 7.50 & 1.01 & 4.23 & 11.92 & 2.56 & 7.83 \\
& GXPF1 & 8.42 & 1.00 & 5.19 & 12.97 & 2.59 & 8.79 \\
& KB3G  & 7.13 & 0.83 & 4.62 & 11.02 & 2.34 & 8.24 \\
& pn-QRPA  & 5.80 & 3.38 & 3.59 & 8.03  & 4.40 & 7.41 \\
& EXP$^{(a)}$  &  3.7  & 2.2 & 3.5  & - & - & - \\
& EXP$^{(b)}$  &  3.0  & 2.3 & 3.1  & 8.3 & 3.6 & 7.5 \\
& EXP$^{(c)}$  &  -  & - & -  & 7.4 & 3.1 & 4.0 \\
& EXP$^{(d)}$  &  -  & - & -  & - & - & 7.8 \\
& EXP$^{(e)}$  &  -  & - & -  & 7.6 & 3.2 & 6.0 \\
\hline $^{55}$Fe & & & & & & &\\
& FPD6  & 4.42 & 1.43 & 3.21 & 14.87 & 3.02 & 8.61 \\
& GXPF1 & 5.07 & 1.26 & 4.52 & 15.93 & 3.24 & 9.92 \\
& KB3G  & 4.03 & 1.08 & 3.82 & 13.96 & 2.95 & 9.22 \\
& pn-QRPA  & 7.12 & 1.75 & 2.98 & 13.68 & 5.77 & 8.31 \\
\hline $^{56}$Fe & & & & & & &\\
& FPD6  & 6.12 & 1.75 & 2.38 & 12.79 & 3.00 & 9.58 \\
& GXPF1 & 6.73 & 1.38 & 3.70 & 14.09 & 3.39 & 10.90 \\
& KB3G  & 5.49 & 1.32 & 3.10 & 12.54 & 3.13 & 10.30 \\
& pn-QRPA  & 3.14 & 1.53 & 2.36 &  7.79 & 3.79 & 9.99 \\
& EXP$^{(a)}$  &  2.7  & 2.0 & 2.3  & - & - & - \\
& EXP$^{(d)}$  &  -  & - & -  & - & - & 9.9 \\
& EXP$^{(f)}$  &  3.5  & 3.2 & 3.2  & - & - & - \\
\end{tabular}
\end{center}
\end{table}

\begin{table}
\caption{\scriptsize Fine-grid calculation of electron capture and
neutrino cooling rates due to $^{54,55,56}$Fe for selected stellar
density of  10$^{11} g/cm^{3}$ as a function of temperature in
stellar matter. log$\rho Y_{e}$ has units of $g/cm^{3}$, where
$\rho$ is the baryon density and $Y_{e}$ is the ratio of the lepton
number to the baryon number. Temperatures ($T_{9}$) are given in
units of $10^{9}$ K. The calculated Fermi energy is denoted by
$E_{f}$ and is given in units of MeV. $\lambda_{ec}$
($\lambda_{\nu}$) are the electron capture (neutrino cooling) rates
in units of $s^{-1}$ ($MeV s^{-1}$). All calculated rates are
tabulated in logarithmic (to base 10) scale.} \label{ta5}
\begin{center}
\scriptsize \begin{tabular}{ccc|cc|cc|cc}  $log\rho Y_{e}$ & $T_{9}$
& $E_{f}$ & \multicolumn{2}{c|}{$^{54}$Fe}&
\multicolumn{2}{|c|}{$^{55}$Fe}
& \multicolumn{2}{|c}{$^{56}$Fe} \\
\cline{4-9} & & & \multicolumn{1}{c}{$\lambda_{ec}$} &
\multicolumn{1}{|c|}{$\lambda_{\nu}$} &
\multicolumn{1}{c}{$\lambda_{ec}$} &
\multicolumn{1}{|c|}{$\lambda_{\nu}$} &
\multicolumn{1}{c}{$\lambda_{ec}$} &
\multicolumn{1}{|c}{$\lambda_{\nu}$} \\ \hline
    11.0 &  0.50 &   23.934  &   4.408    &    5.593  &  4.355   &  5.493  &  4.255   &    5.389  \\
    11.0 &  1.00 &   23.933  &   4.408   &     5.593 &   4.355  &   5.493 &   4.255  &     5.389   \\
    11.0 &  1.50 &   23.932  &   4.409   &     5.594 &   4.355  &   5.492 &   4.255  &     5.390 \\
    11.0 &  2.00 &   23.930  &   4.409   &     5.594 &   4.355  &   5.492 &   4.255  &     5.390   \\
    11.0 &  2.50 &   23.928  &   4.409   &     5.594 &   4.355  &   5.493 &   4.256  &     5.390 \\
    11.0 &  3.00 &   23.925  &   4.409   &     5.595 &   4.356  &   5.494 &   4.256  &     5.390   \\
    11.0 &  3.50 &   23.922  &   4.410   &     5.596 &   4.357  &   5.496 &   4.256  &     5.390  \\
    11.0 &  4.00 &   23.918  &   4.411   &     5.596 &   4.358  &   5.499 &   4.257  &     5.390  \\
    11.0 &  4.50 &   23.913  &   4.411   &     5.597 &   4.360  &   5.502 &   4.257  &     5.390   \\
    11.0 &  5.00 &   23.908  &   4.412   &     5.598 &   4.362  &   5.505 &   4.258  &     5.390    \\
    11.0 &  5.50 &   23.903  &   4.413   &     5.600 &   4.364  &   5.508 &   4.259   &    5.391   \\
    11.0 &  6.00 &   23.897  &   4.414   &     5.601 &   4.367   &  5.512 &   4.260    &   5.392  \\
    11.0 &  6.50 &   23.891  &   4.416   &     5.603 &   4.369  &   5.516 &   4.261   &    5.393  \\
    11.0 &  7.00 &   23.884  &   4.417   &     5.605 &   4.372  &   5.521 &   4.262   &    5.395  \\
    11.0 &  7.50 &   23.877  &   4.419   &     5.607 &   4.376  &   5.526 &   4.264   &    5.397    \\
    11.0 &  8.00 &   23.869  &   4.421   &     5.610 &   4.380  &   5.532 &   4.266   &    5.399   \\
    11.0 &  8.50 &   23.860  &   4.423   &     5.613 &   4.384  &   5.538 &   4.269   &    5.402    \\
    11.0 &  9.00 &   23.851  &   4.426   &     5.617 &   4.389  &   5.545 &   4.272   &    5.406 \\
    11.0 &  9.50 &   23.842  &   4.429   &     5.621 &   4.395  &   5.552 &   4.275   &    5.410   \\
    11.0 &  10.00 &  23.832  &   4.432   &     5.626 &   4.401  &   5.560 &   4.278   &    5.415  \\
    11.0 &  15.00 &  23.704  &   4.481   &     5.696 &   4.481  &   5.666  &  4.335   &    5.492 \\
    11.0 &  20.00 &  23.526  &   4.543   &     5.784 &   4.568  &   5.787  &  4.406   &    5.594  \\
    11.0 &  25.00 &  23.296  &   4.599   &     5.865 &   4.645  &   5.900  &  4.471   &    5.689   \\
    11.0 &  30.00 &  23.016  &   4.648   &     5.937 &   4.712  &   6.000  &  4.527   &    5.773   \\
\end{tabular}
\end{center}
\end{table}
\end{document}